\documentclass[twocolumn,pra,aps,amsfonts]{revtex4}
\usepackage{amsmath}
\usepackage{graphicx}
\usepackage{color}

\newcommand{\<}{\langle}
\renewcommand{\>}{\rangle}
\newcommand{\ud}{\mathrm{d}}
\DeclareMathOperator{\arctanh}{arctanh}
\DeclareMathOperator{\tr}{Tr}
\newcommand{\Sga}{S_\alpha(\gamma_A)}
\newcommand{\piha}{\frac{\pi}{2}}

\newcommand{\PZ}{{\cal P}^\emptyset}

\newtheorem{definition}{Definition}
\newtheorem{Fact}{Fact}

\begin{document}

\author{\L{}ukasz Pankowski $^{(1)}$ and Barbara Synak-Radtke $^{(2)}$}
\title{Can quantum correlations be completely quantum?}

\affiliation{$^{(1)}$Department of Mathematics Physics and Computer
  Science, University of Gda\'nsk, Poland}
\affiliation{$^{(2)}$Institute of Theoretical Physics and
  Astrophysics, University of Gda\'nsk, Poland}

\begin{abstract}
  Deficit of information zero-way was proposed in \cite{huge-delta} as
  one of possible measures of quantumness of correlations.  Numerical
  calculations suggested that there exist such states for which this
  quantity is almost equal to mutual information.  In this paper we
  present a family of states for which we have equality between above
  measure of quantumness of correlations and the measure of total
  correlations --- mutual information.  It means that whole
  correlations in these states have, in some sense, quantum character
  and that quantum correlations do not necessarily imply classical
  correlations.  We prove this intriguing feature for a subclass of
  $2\otimes2$ separable states.  We also present numerical result
  suggesting that this interesting situation might also happen for
  $2\otimes2$ entangled states.
\end{abstract}

\maketitle

\section{Introduction}

It has been found in different contexts
\cite{Bennett-nlwe,OlliverZ2001,Zurek2003,OHHH2001,huge-delta} that
entanglement does not exhaust quantumness of correlations contained in
compound quantum systems.  In order to characterize quantumness of
states the \emph{properly classically correlated} bipartite states
(shortly classically correlated) were defined \cite{OHHH2001}.  These
states can be written in the form
\begin{align}
  \label{eq:classically-correlated}
  \rho_{AB} = \sum_{ij} p_{ij}|i\>\<i| \otimes |j\>\<j|
\end{align}
with coefficients $0 \leq p_{ij}\leq 1$, $\sum_{ij} p_{ij} =1$;\,
$\{|i\>\}$ and $\{|j\>\}$ are local bases.  The set of classically
correlated states is invariant under local unitary operations.  These
states are diagonal in the so called biproduct basis.  The
\emph{non-classically} correlated states are those that cannot be
represented in the above form.

In \cite{OHHH2001} a measure of quantumness of correlations so called
quantum deficit $\Delta$ was introduced, which is zero for classically
correlated states (see extensive development \cite{huge-delta}).  In
particular, it has been shown that there exist separable states which
have nonzero deficit.  This means that those separable states exhibit
classical correlations between quantum properties.  The paper
\cite{huge-delta} introduces variants of quantum deficit with
restricted communication as independent candidates for the measure of
quantumness of correlations.  One of them called quantum deficit
zero-way $\Delta^\emptyset$ is equivalent to the distance from
classically correlated states \cite{huge-delta}.  Recently similar
measures of quantumness of correlations based on the distance from
classically correlated states has been introduced \cite{gkm07,srn07}
and quantumness of some subclasses of separable and entangled states
was investigated.

The main purpose of the present paper is to inquire a relation between
the measure of total correlations --- mutual information and quantum
deficit.  The later is defined \cite{OHHH2001} as the difference
between informational content of a state and information that can be
localized to a subsystem by use of local unitary operations and a
dephasing channel.  Quantum deficit refers us to this part of
correlations that must be destroyed during the process of localizing
information into a subsystem.  In this paper we consider a special
kind of quantum deficit called zero-way ($\Delta^\emptyset$)
\cite{huge-delta} where only a restricted class, called $\PZ$, of
protocols of localizing information is permitted.  In $\PZ$ protocols
we make local measurements and only after this we use a classical
channel to collect data and then exploit classical correlations
created by the measurements to localize information.  Surprisingly, we
find such states for which quantum deficit zero-way is equal to mutual
information $I_M$ (the measure of total correlations in a state),
i.e. $\Delta^\emptyset = I_M$.  It means that all correlations in
these states have quantum character.  Notice that for maximally
entangled states $\Delta^\emptyset= \frac{1}{2} I_M$, so we can say
that only half of correlations manifest quantumness.  In this context
the $\Delta^\emptyset=I_M$ feature of the states presented in this
paper seems especially interesting.

The notion of quantum deficit is built by use of quantities which has
operational meaning in the regime of many copies: information $I$ is
the number of pure states we can distill from a given state by so
called Noisy Operations; localizable information $I_l$ is equal to the
number of pure product states we can distill by CLOCC operation in the
asymptotic regime.  These two quantities can also be interpreted in
terms of work we can draw from a heat bath by use of a given quantum
state.  Quantum deficit in the asymptotic regime inherits this clean
operational meaning.  But, on the other hand, it is much harder to
evaluate.  Some bounds on the regularized quantum deficit have been
obtained in \cite{SynakHH2005}.  In our case we consider one copy
version of quantum deficit and additionally we allow only for a
special kind of protocols of localizing information into a subsystem.
In spite of such strong restrictions we get quite interesting result.

\section{Basic notion}

{\it Informational content $I$ of a state.}  Information is an
abstract concept.  Here we will use this term to refer to a specific
function $I$
\begin{align}
  I(\varrho) = N - S(\varrho)
\label{eq:infofunc}
\end{align}
where $\varrho$ is a state of $N$ qubits, and
$S(\varrho)=-\tr\varrho\log\varrho$ is the von Neumann entropy.  The
information $I$ has operational meaning in the asymptotic regime of
many identical copies\cite{nlocc}.  This is the unique function (up to
constants) that is not increasing under the class of so called Noisy
Operations (NO) \cite{uniqueinfo}: operations that consist of (i)
unitary transformations, (ii) partial trace, and (iii) adding ancilla
in maximally mixed state.  One can then show that $I$ determines the
optimal rate of transitions between states under NO.

{\it By CLOCC operations} on bipartite system of $n_{AB}$ qubits we
mean all operations that can be composed out of
\begin{itemize}
\item[(i)] local unitary transformations,
\item [(ii)] sending a subsystem down completely decohering
  (dephasing) channel.
\end{itemize}
Notice that CLOCC class is a subclass of LOCC operations (Local
Operations and Classical Communication) and is equivalent to local
measurements, local unitary operations and classical communication.
The restrictions of CLOCC (in contrast to LOCC we are forbidden to add
pure ancillas and discard a local subsystem) are a consequence of the
fact that we must control information flow.  Every state (besides
maximally mixed ones) has nonzero value of $I$, so especially pure
ancillas.

{\it Localizable information} $I_l(\varrho_{AB})$ of a state
$\varrho_{AB}$ is the maximal amount of local information that can be
obtained by CLOCC operations.
\begin{align}
  I_l(\varrho_{AB}) =
  \sup_{\Lambda \in CLOCC}(I(\varrho'_A) + I(\varrho'_B))
\end{align}
where $\varrho'_{AB} =\Lambda (\varrho_{AB})$.

{\it Local information $I_{LO}$} is the difference between the number
of qubits $N$ of a state and the sum of entropies of its subsystems
\begin{align}
  I_{LO}(\varrho_{AB}) = N - S_A(\varrho_{AB}) - S_B(\varrho_{AB})
\end{align}

{\it Mutual information $I_M$} of a state $\varrho_{AB}$ is the
difference between the sum of local entropies and the entropy of the
state.
\begin{align}
  I_M(\varrho_{AB}) =
  S_A(\varrho_{AB}) + S_B(\varrho_{AB}) - S(\varrho_{AB})
\end{align}

\section{Quantum and classical deficit zero way}

In this section we introduce the quantities which are fundamental for
our consideration: quantum deficit and classical deficit.

\begin{definition}
  The quantum deficit $\Delta (\varrho_{AB})$ of a state
  $\varrho_{AB}$ is given by the difference between informational
  content $I$ of the state and localizable information $I_l$
  \begin{align}
    \Delta(\varrho_{AB}) = I(\varrho_{AB}) - I_l(\varrho_{AB})
  \end{align}
\end{definition}
Notice that quantum deficit can be rewritten as
\begin{align}
  \Delta(\varrho_{AB}) =
  \inf_{\Lambda\in CLOCC}S(\varrho'_{AB}) - S(\varrho_{AB})
\end{align}
where $\varrho'_{AB} =\Lambda(\varrho_{AB})$ and $\Lambda$ is
optimized over CLOCC operations.

Quantum deficit tells us about the amount of information that cannot
be localized into a subsystem.  It means that part of information is
necessarily destroyed in the process of localizing information by use
of a classical channel.  So this part must be somehow quantum and come
from correlations.  This is the reason why we interpret quantum
deficit as a measure of quantumness of correlations.
\begin{definition}
  The classical deficit $\Delta_{cl}(\varrho_{AB})$ of a quantum state
  is the difference between the information that can be localized by
  means of CLOCC operations (i.e. localizable information $I_l$) and
  local information
  \begin{align}
    \Delta_{cl}(\varrho_{AB}) = I_l(\varrho_{AB}) - I_{LO}(\varrho_{AB})
  \end{align}
\end{definition}
Classical deficit tells us how much more information can be obtained
from a state $\varrho_{AB}$ by exploiting additional correlations in
the state using a classical channel.

Notice that quantum and classical deficit add up to mutual information
$I_M$.
\begin{align}
  \Delta + \Delta_{cl} = I_M
\end{align}
Thus we can express classical deficit as follows
\begin{align}
  \Delta_{cl} = I_M - \Delta
\end{align}

We can restrict classical communication between Alice and Bob to
one-way communication (from Alice to Bob or from Bob to Alice) or to
so called zero-way communication.

Consider zero-way protocol $\PZ$ of localizing information consisting
of complete local measurement (or local complete dephasing) and
classical communication which is allowed to be performed only after
making measurement (or dephasing).  Classical communication is
necessary to collect data and exploit the pure classical correlations
in order to localize information.  Notice that the main difference
between a zero-way protocol and a general CLOCC protocol of localizing
information is that we are forbidden to communicate classically before
all measurement or dephasing operations are finished.  This guaranties
that we are not able to draw any nonclassical information from
correlations in a state, for example, from correlations of subsystems
of some separable states which we can call classical correlations
between quantum properties of the state.

If we restrict classical communication in localizable information
$I_l$ to zero-way protocols $\PZ$ we get localizable information
zero-way $I_l^\emptyset$
\begin{align}
  I_l^\emptyset(\varrho_{AB}) =
  \sup_{\Lambda \in \PZ}(I(\varrho'_A) + I(\varrho'_B))
\end{align}

Now we can define quantum deficit zero-way $\Delta^\emptyset$ and
classical deficit zero-way $\Delta_{cl}^\emptyset $.

\begin{definition}
  The quantum deficit zero-way $\Delta^\emptyset (\varrho_{AB})$ of a
  state $\varrho_{AB}$ is given by the difference between
  informational content $I$ of the state and information
  $I_l^\emptyset$ localizable by zero-way protocol $\PZ$
\begin{align}
  \Delta^\emptyset(\varrho_{AB}) =
  I(\varrho_{AB}) - I_l^\emptyset(\varrho_{AB})
\end{align}
\end{definition}
Equivalently, we can express quantum deficit zero-way as
\begin{align}
  \label{eq:Delta0-equiv}
  \Delta^\emptyset(\varrho_{AB}) =
  \inf_{\Lambda \in \PZ}S(\varrho'_{AB}) - S(\varrho_{AB})
\end{align}
where $\varrho'_{AB}=\Lambda(\varrho_{AB})$ and $\Lambda$ is optimized
over $\PZ$ protocols.  One can also find that quantum deficit zero-way
$\Delta^\emptyset$ is equal to relative entropy distance from the set
of classically correlated states \cite{huge-delta}.

The quantum deficit zero-way is an independent candidate for measure
of quantumness of correlations.  States having no quantum correlations
should have $\Delta^\emptyset =0$.  Such restricted measure can
capture interesting aspects of nonlocality.  Consider for example a
bipartite state with eigenbasis of the form
\begin{multline}
  \label{onewaydistin}
  |0\>_A |0\>_B, |0\>_A |1\>_B, \\
  \frac{1}{\sqrt{2}} |1\>_A (|0\>+ |1\>)_B,
  \frac{1}{\sqrt{2}} |1\>_A (|0\>- |1\>)_B
\end{multline}
such basis is locally indistinguishable and is also not
distinguishable by zero-way communication.  Therefore a mixture of the
states (\ref{onewaydistin}) where the mixing probabilities are all
different from each other would have nonvanishing $\Delta ^\emptyset$
\cite{huge-delta}.  This is in contrast to states which are mixtures
of the set of states
\begin{displaymath}
  |0\>_A |0\>_B, |0\>_A |1\>_B, |1\>_A |0\>_B, |1\>_A |1\>_B
\end{displaymath}
for which all the information is extractable from the state locally,
by measurement by both the parties without any communication.  It
suggests that the quantum behaviour of correlations could result from
distinctly quantum but ``local'' properties of nonorthogonality.  This
is connected with examples of LOCC-indistinguishability of orthogonal
product basis \cite{Bennett-nlwe,BennettUPBI1999}.

In analogy to the quantum deficit zero-way $\Delta^\emptyset$ we can
define the classical deficit zero-way $\Delta_{cl}^\emptyset$.

\begin{definition}
  The classical deficit zero-way $\Delta_{cl}^\emptyset(\varrho_{AB})$
  of a state $\varrho_{AB}$ is the difference between information
  $I_l^\emptyset$ localizable by zero-way protocol $\PZ$ and local
  information
  \begin{align}
    \Delta_{cl}^\emptyset(\varrho_{AB}) =
    I_l^\emptyset(\varrho_{AB}) - I_{LO}(\varrho_{AB})
  \end{align}
\end{definition}
or equivalently we have that
\begin{align}
  \Delta_{cl}^\emptyset = I_M - \Delta^\emptyset
\end{align}
Classical deficit zero-way tells us about the amount of information
that can be localized into a subsystem from correlations which are not
destroyed after local measurement (or dephasing).  Notice that after
measurement a state changes into classically correlated one
\eqref{eq:classically-correlated} and then all information present in
correlations of such a state can be localized.

There is a question how great can be quantum deficit.  We know that
for a pure state $\psi$ it is given by
\begin{align}
  \Delta(\psi) = \frac{1}{2} I_M(\psi)
\end{align}

So in particular for maximally entangled state $\psi^+$ we have
$\Delta (\psi^+) =\frac{1}{2}I_M(\psi^+)$.  Additionally for pure
states $\Delta^\emptyset= \Delta $, because the greatest value of
localizable information is equal to the information which can be
concentrated to a subsystem using $\PZ$ protocol.  We can ask if there
exist mixed states for which the rate of quantum deficit to $I_M$ is
greater than for maximally entangled states, so if there are states
for which amount of quantumness of correlations is greater than
$\frac{1}{2} I_M$.  We are able to answer this question for quantum
deficit zero-way.  What is more we can find such states for which
$\Delta^\emptyset=I_M$.  It implies immediately that for these states
$\Delta_{cl}^\emptyset=0$.  This means that after optimal local
measurement (or dephasing) all correlations are destroyed, so by $\PZ$
protocol we cannot extract any additional information than in local
scenario.

\begin{figure}
  \centering
  \input{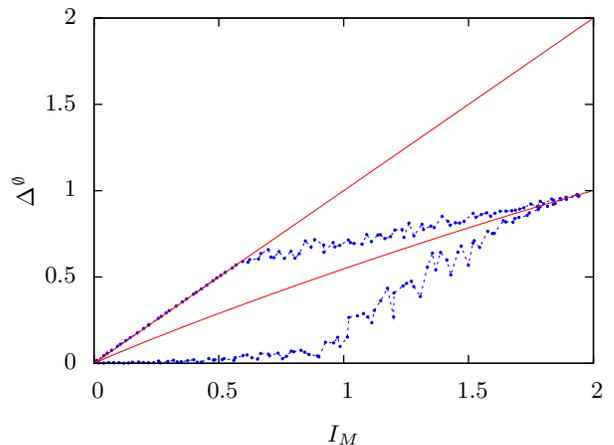}
  \caption{Deficit zero-way is plotted versus mutual information for
    100 000 random two qubit states, only maximal and minimal values
    of $\Delta^\emptyset$ are plotted in short intervals of $I_M$.
    The upper line is an upper bound for $\Delta^\emptyset$, while the
    lower one is $\Delta^\emptyset$ of isotropic states.}
  \label{fig:delta0-vs-im}
\end{figure}

On figure \ref{fig:delta0-vs-im} we can see that there exist states
for which $\Delta^\emptyset$ is equal or almost equal to $I_M$.  So
there is a task to find such states for which this equality holds.

\section{States for which $\Delta^\emptyset=I_M$}

In this section we introduce a class of states $\varrho_a$
parameterized with $a$ and prove that these states up to some value of
parameter $a$ fulfill the condition that
$\Delta^\emptyset(\varrho_a)=I_M(\varrho_a)$.

\subsection{Defining the class of states}

Let us define the class of states $\varrho_a$ for $a\in[0,1]$ on
$2\otimes2$ Hilbert space as
\begin{align}
  \label{eq:rho_a}
  \varrho_a =
  \frac{1}{2} \left( |\phi_a\>\<\phi_a| + |\psi_a\>\<\psi_a| \right)
\end{align}
where
\begin{align}
  \label{eq:phi_a}
  |\phi_a\> &= \sqrt{a} \, |00\> + \sqrt{1 - a} \, |11\> \\
  \label{eq:psi_a}
  |\psi_a\> &= \sqrt{a} \, |01\> + \sqrt{1 - a} \, |10\>
\end{align}
In our problem $\varrho_a$ and $\varrho_{1-a}$ are equivalent (they
only differ by a local operation) thus we will only consider
$a\in[0,\frac12]$.

The mutual information for states $\varrho_a$ is given by
\begin{align}
  I_M(\varrho_a) = H(a)
\end{align}
thus the value of $I_M$ may be freely chosen in the interval $[0, 1]$
by proper selection of parameter $a$.

\begin{figure}[tbp]
  \centering
  \input{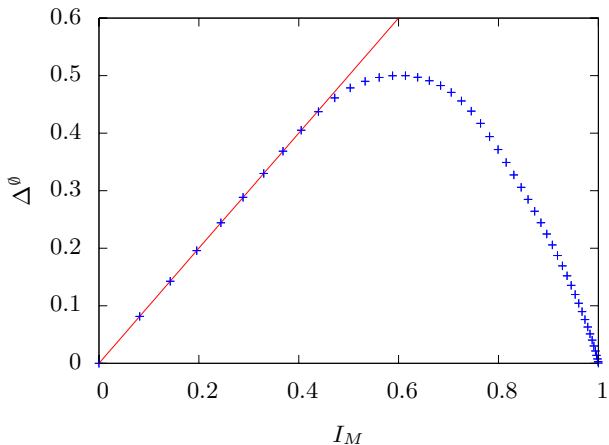}
  \caption{Deficit zero-way is plotted versus mutual information for
    the class of states $\varrho_a$.  The line is an upper bound for
    $\Delta^\emptyset$.}
  \label{fig:rho-a-experim}
\end{figure}

The deficit zero-way for states $\varrho_a$ is given by
\begin{align}
  \label{eq:delta0-varho_a}
  \Delta^\emptyset(\varrho_a)
  = \inf_{\Lambda\in\PZ} S(\varrho_a') - 1
\end{align}

Numerical deficit zero-way versus mutual information for the states of
our class $\varrho_a$ suggests (figure~\ref{fig:rho-a-experim}) that
for some interval $[0,a_0]$ of parameter $a$ we have equality between
quantities $\Delta^\emptyset$ and $I_M$.

\subsection{Sketch of the proof}

To optimize the value of $\Delta^\emptyset$ given by
\eqref{eq:delta0-varho_a} we start from the formula for $S(\varrho')$
reachable under $\PZ$ protocols.  Then we reduce the optimization of
$S(\varrho')$ to the optimization of single real parameter function
$\Sga$, where $\alpha$ is determined by $a$.  Later by analyzing the
first and second order derivatives of $\Sga$ we argue that for
$\alpha\in[\alpha_0,\piha]$ corresponding to $a \in [0,a_0]$ there is
the global minimum of $\Sga$ in $\gamma_A=0$ which implies
$\Delta^\emptyset=I_M$.

On the other hand for $\alpha\in [0,\alpha_0)$ corresponding to $a \in
(a_0,\frac12]$ there is a maximum of $\Sga$ in $\gamma_A=0$ which
implies $\Delta^\emptyset < I_M$.  Finally we compute the value of
$a_0$.

\subsection{Helpful functions}

First we introduce two functions, which will be useful in the further
consideration.  Let us define
\begin{align}
  H_s(x) &= H\left(\frac{1+\sin(x)}{2}\right) \\
  H_c(x) &= H\left(\frac{1+\cos(x)}{2}\right)
\end{align}

The above functions have the following properties
\begin{enumerate}
\item $H_s$ and $H_c$ are even so their first derivatives are odd and
  the second are even.
\item $H_s$ and $H_c$ are periodic with period $\pi$ and so are
  derivatives.
\item $H_s(\frac{\pi}{2} \pm x) = H_c(x)$ and similar for derivatives.
\end{enumerate}

\subsection{Simplifying the form of $S(\varrho'_a)$}

For any two qubit state $\varrho'_a$ reachable under $\PZ$ protocol
its von Neumann entropy can be expressed by
\begin{align}
  \label{eq:Sp-diagonal}
  S(\varrho_a') = H(\mathrm{diagonal}(
  U_A \otimes U_B \; \varrho_a \; U_A^\dagger \otimes U_B^\dagger))
\end{align}
where $U_A$ and $U_B$ are single qubit unitaries which can be
parameterized as follows \cite{Nielsen-Chuang}:
\begin{align}
  U &= e^{i\alpha}
  \begin{bmatrix}
    e^{i\*\left(-\frac{\beta}{2}-\frac{\delta}{2}\right)}
    \*\cos\left(\frac{\gamma}{2}\right) &
    -e^{i\*\left(-\frac{\beta}{2} + \frac{\delta}{2}\right)}
    \*\sin \left(\frac{\gamma}{2}\right)\\[0.2ex]
    e^{i \*\left(\frac{\beta}{2}-\frac{\delta}{2}\right)}
    \*\sin \left(\frac{\gamma}{2} \right) &
    e^{i\*\left(\frac{\beta}{2}+\frac{\delta}{2}\right)}
    \*\cos \left(\frac{\gamma}{2}\right)
  \end{bmatrix}
\end{align}

Using this parameterization for $U_A$ and $U_B$ we can simplify with
help of Maxima\footnote{Maxima (http://maxima.sourceforge.net/) is a
  free software computer algebra system released under GNU GPL.} the
form of $S(\varrho'_a)$ to
\begin{align}
  \label{eq:S'-distrib}
    S(\varrho'_a) = \textstyle H\left(
      \frac{1 + s - c}{4},
      \frac{1 - s - c}{4},
      \frac{1 - s + c}{4},
      \frac{1 + s + c}{4}
    \right)
\end{align}
where
\begin{align}
  \label{eq:s}
  s &= 2\*\sqrt{a(1-a)}
  \sin\gamma_A \sin\gamma_B
  \cos\delta_A \cos\delta_B \\
  c &= \left(1-2\*a\right)\*\cos \gamma_A
\end{align}
(We subscript the parameters with the name of the subsystem).  Note
that this form of $S(\varrho'_a)$ exhibits asymmetry with respect to
parameters of $U_A$ and $U_B$.  Our class of states $\varrho_a$ is
indeed asymmetric (i.e. $\varrho_a \neq V \varrho_a V$ where $V$ is
the swap operator) unless $a=\frac{1}{2}$, for example,
$S_A(\varrho_a) \neq S_B(\varrho_a)$.  The asymmetry with respect to
parameters of $U_A$ and $U_B$ will appear even stronger after the next
simplification.

Let us here recall the fact:
\begin{Fact}
  \label{fact:prob-average-increase-H}
  Any change toward equalization of probabilities $p_1, p_2, \ldots,
  p_n$ increases $H$.  Thus if $p_1 < p_2$ and we increase $p_1$,
  decreasing $p_2$ an equal amount so that $p_1$ and $p_2$ are more
  nearly equal, then $H$ increases. (\ldots) \cite{Shannon1948}
\end{Fact}

We group probabilities from distribution \eqref{eq:S'-distrib} in two
inequalities
\begin{align}
  \frac{(1 \pm c) - |s|}{4} \le \frac{(1 \pm c) + |s|}{4}
\end{align}
Now we can see that decreasing $|s|$ for any fixed value of $c$ will
change both inequalities towards equalization and hence from fact
\ref{fact:prob-average-increase-H} will increase entropy
\eqref{eq:S'-distrib}.  As we are minimizing $S(\varrho'_a)$ in
\eqref{eq:delta0-varho_a} thus we should maximize $|s|$ in respect to
the parameters with no influence on $c$.  We can do this by setting
\begin{align}
  \sin\gamma_B \cos\delta_A \cos\delta_B  = 1
\end{align}
in \eqref{eq:s}.  After this substitution we observe that $s + c$ and
$s - c$ are harmonic oscillations, i.e.
\begin{align}
  \label{eq:s-pm-c-simp}
  s \pm c = \sin(\gamma_A \pm \alpha)
\end{align}
where $\alpha$ is determined by
\begin{align}
  \sin \alpha &= 1 - 2 a \\
  \cos \alpha &= 2 \sqrt{a(1-a)}
\end{align}

The harmonic oscillations of \eqref{eq:s-pm-c-simp} allow us to
simplify $S(\varrho'_a)$ given by \eqref{eq:S'-distrib} to
\begin{align}
  S_\alpha(\gamma_A) =
  1 + \frac{1}{2} H_s (\gamma_A + \alpha) + \frac{1}{2} H_s(\gamma_A - \alpha)
\end{align}
with the property that
\begin{align}
  \inf_{\Lambda\in\PZ} S(\varrho_a') =
  \inf_{\gamma_A\in[-\piha,\piha]} S_\alpha(\gamma_A)
\end{align}

\subsection{Proving $\Delta^\emptyset=I_M$ for $a\in[0,a_0]$}

First we observe that
\begin{align}
  S_\alpha(0) = 1 + H(a)
\end{align}
thus if $\gamma_A=0$ is the global minimum of $\Sga$ for some $\alpha$
we get the desired equality
\begin{align}
  \Delta^\emptyset(\varrho_a) = (1 + H(a)) - 1 = I_M(\varrho_a)
\end{align}
On the other hand if $\gamma_A=0$ is a maximum of $\Sga$ for some
$\alpha$ than the value of $\Sga$ in the global minimum is less than
$S_\alpha(0)$ and thus $\Delta^\emptyset < I_M$.

\begin{figure}
  \centering
  \input{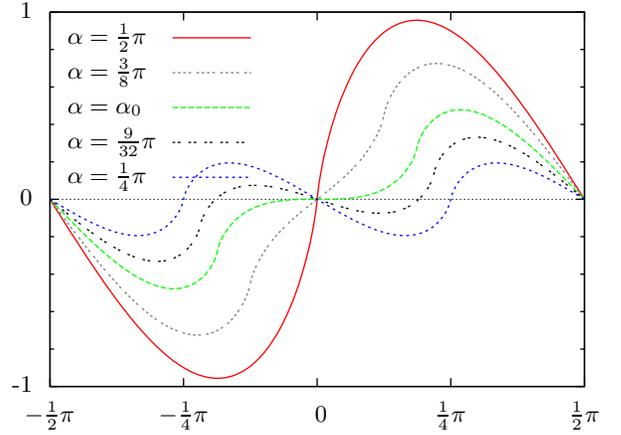}
  \caption{First order derivative of $\Sga$ for sample values of
    $\alpha$ illustrating the appearance of two more zeros.}
  \label{fig:s-ga-deriv}
\end{figure}

In search for extrema of $\Sga$ we consider its derivative
\begin{align}
  \label{eq:diff-Sga}
  \frac{\ud S_\alpha(\gamma_A)}{\ud \gamma_A}
  = \frac{1}{2} \left(
    \frac{\ud H_s(\gamma_A + \alpha)}{\ud \gamma_A}
    + \frac{\ud H_s(\gamma_A - \alpha)}{\ud \gamma_A}
  \right)
\end{align}

Notice that $\Sga$, same as $H_s$, is periodic with period $\pi$.
Thus it is enough to consider the range of a single period $\gamma_A
\in [-\piha,\piha]$.  We identify both ends of the period when it
comes to listing zeros in a period.  In the interval $[-\piha,\piha]$
the derivative of $\Sga$ has for all $\alpha$ zeros in $\gamma_A=0$
and $\gamma_A=\frac{\pi}{2}$ (as derivatives of $H_s$ and $H_c$ are
odd; figure \ref{fig:s-ga-deriv}).

We also consider the second order derivative of $\Sga$ which is the
average of two \emph{impulses} relatively shifted by $2\alpha$.  By
the \emph{impulse} we mean a function of the form
\begin{align}
  \frac{\ud^2 H_s(x)}{\ud x^2} =
  \frac{1}{2} \sin(x) \cdot (\log \circ \, f \circ \sin)(x) - \frac{1}{\ln 2}
\end{align}
where
\begin{align}
  f(x) = \frac{1 + x}{1 - x}
\end{align}

\begin{figure}
  \centering
  \input{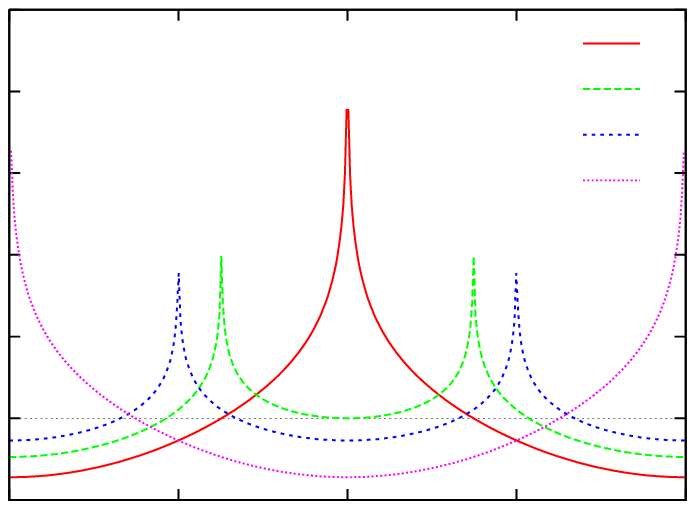}
  \caption{Second order derivative of $\Sga$ for the most
    characteristic values of the range $\alpha\in[0,\piha]$.}
  \label{fig:s-ga-second-deriv}
\end{figure}

Since $f$ is strictly increasing in the interval $[-1,1)$, thus $f
\circ \sin$ is strictly increasing in $[0,\frac{\pi}{2})$ and strictly
decreasing in $(-\frac{\pi}{2}, 0]$ and the same holds for $\log \circ
\, f \circ \sin$ and the impulse.  The impulse has a \emph{peak} in
$\piha$ by which we mean it tends to infinity in this point.  The
impulse also has one negative and one positive interval per period.
It can be observed on figure \ref{fig:s-ga-second-deriv} as the curve
labeled $\alpha=0$.

Now we analyze the extrema of $\Sga$ for $\alpha\in[0,\piha]$, which
corresponds to $a\in [0,\frac12]$.

For $\alpha=\piha$ corresponding to $a=0$ both impulses of the second
order derivative of $\Sga$ are equal and have peaks in $\gamma_A=0$
(figure \ref{fig:s-ga-second-deriv}) and thus $\gamma_A=0$ is a
minimum of $\Sga$.  If we move with $\alpha$ from $\piha$ down to 0
than peaks of both impulses are shifting away from $\gamma_A=0$ and
thus the value of the second order derivative of $\Sga$ in
$\gamma_A=0$ is strictly monotonically decreasing to $(-\ln2)^{-1}$
for $\alpha=0$.  The point in which it reaches 0 on this way will be
called $\alpha_0$.

Thus for $\alpha\in[0, \alpha_0)$ corresponding to $a\in(a_0,
\frac12]$ the second order derivative is negative in $\gamma_A=0$ and
so $\Sga$ has maximum in $\gamma_A=0$ which implies $\Delta^\emptyset
< I_M$.

For $\alpha\in[\alpha_0,\piha]$ corresponding to $a\in[0,a_0]$ the
second order derivative of $\Sga$ has one negative and one positive
interval per period which implies that the first order derivative has
only one strictly increasing and one strictly decreasing interval per
period and thus at most two zeros (one per interval).  So it has
exactly two zeros: $\gamma_A=0$ and $\gamma_A=\piha$, which must
appear for all $\alpha$.  For $\alpha\in[\alpha_0,\piha]$ the first
order derivative of $\Sga$ is negative in $\gamma_A=\piha$ thus
$\gamma_A=\piha$ is the global maximum of $\Sga$ which implies
$\gamma_A=0$ is the global minimum and so $\Delta^\emptyset=I_M$ for
$a\in[0,a_0]$.

The last step is to find $a_0$.

\subsection{Finding $a_0$}

To find $a_0$ we numerically solve the equation
\begin{align}
  \left. \frac{\ud^2 \Sga}{\ud \gamma_A^2} \right|_{\gamma_A = 0} &=
  (1 - 2a) \arctanh (1 - 2a)  - 1 = 0
\end{align}
and the smaller of two roots gives $a_0$
\begin{align}
  a_0 \approx 0.08322
\end{align}

\section{Reaching lower bound on $\Delta^\emptyset$ for a given $I_M$}

On $2\otimes2$ Hilbert space $\Delta_{cl}^\emptyset \le 1$ thus on
this space the lower bound on $\Delta^\emptyset$ is a function of
$I_M$
\begin{align}
  \Delta^\emptyset \ge \max(0, I_M - 1)
\end{align}
We show that for any $I_M$ this lower bound is achievable.

For $I_M \in [0,1]$ the lower bound is achieved by the simple class of
separable states
\begin{align}
  \sigma_p = p |00\>\<00| + (1 - p) |11\>\<11|
\end{align}

While for $I_M \in [1, 2]$ the lower bound is achieved by the class of
states
\begin{align}
  \varrho_p = p\,|\phi^+\>\<\phi^+| + (1 - p)\,|\psi^+\>\<\psi^+|
\end{align}
where
\begin{align}
  |\phi^+\> &= \frac{1}{\sqrt{2}} ( |00\> + |11\> ) \\
  |\psi^+\> &= \frac{1}{\sqrt{2}} ( |01\> + |10\> )
\end{align}

For $\sigma_p$ the infimum of $\Delta^\emptyset$ as given by
\eqref{eq:Delta0-equiv} is achieved by setting $U_A=U_B=I$ in
\eqref{eq:Sp-diagonal} while for $\varrho_p$ by $U_A=U_B=H$, where $H$
is the Hadamard gate.

\section{Another interesting class}

Let us introduce another class of states on $2\otimes2$ Hilbert space
\begin{align}
  \varrho_{a, b, p} = p\,|\phi_a\>\<\phi_a| + (1 - p)\,|\psi_b\>\<\psi_b|
\end{align}
where $|\phi_a\>$ and $|\psi_b\>$ are given by \eqref{eq:phi_a} and
\eqref{eq:psi_a} respectively.  This class is a generalization of two
of previously considered classes, i.e.
\begin{align}
  \varrho_a &= \varrho_{a, a, \frac{1}{2}} \\
  \varrho_p &= \varrho_{\frac{1}{2},\frac{1}{2},p}
\end{align}

\begin{figure}[tbp]
  \centering
  \input{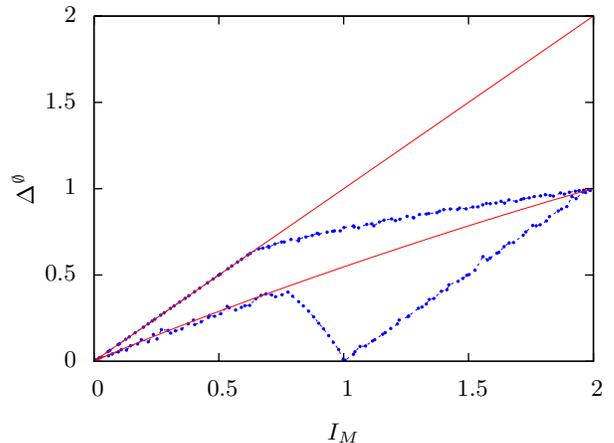}
  \caption{Deficit zero-way is plotted versus mutual information for
    10~000 random two qubit states from $\varrho_{a, b, p}$ class,
    only maximal and minimal values of $\Delta^\emptyset$ are plotted
    in short intervals of $I_M$.  The upper line is an upper bound for
    $\Delta^\emptyset$, while the lower one is $\Delta^\emptyset$ of
    isotropic states.}
  \label{fig:rho_abp}
\end{figure}

Random elements of the $\varrho_{a, b, p}$ class
(figure~\ref{fig:rho_abp}) cover wider range of $\Delta^\emptyset$
than general two qubit states (figure~\ref{fig:delta0-vs-im}), except
some evident region, although there are 100~000 states on
figure~\ref{fig:delta0-vs-im} and only 10~000 on
figure~\ref{fig:rho_abp}.

What is interesting, for $\varrho_{a,b,p}$ states the condition
$\Delta^\emptyset=I_M$ require $\varrho_{a,b,p}$ to be a separable
state.  It comes from the fact that if we want to fulfill this
condition then after measurement made in eigenbasis of subsystems of a
state we have to obtain a product state.  Otherwise we would be able
to get more than $2-S_A-S_B$ localizable information zero-way, because
$S(\varrho') <S_A + S_B$ for nonproduct states.  The above condition for
states $\varrho_{a,b,p}$ can be expressed by the parameters $a$, $b$,
$p$ as follows
\begin{align}
  p^2 a (a - 1) - (1 - p)^2 b (1 - b) = 0
\end{align}
which is equivalent to
\begin{align}
  p \sqrt{a (a - 1)} = (1 - p) \sqrt{b (1 - b)}
  \label{eq:cond}
\end{align}
And equality \eqref{eq:cond} implies via partial transposition
criterion that $\varrho_{a,b,p}$ is a separable state.

\section{Entangled states for which $\Delta^{\emptyset}=I_M$ may hold}

Consider the following class of states (mixtures of two nonorthogonal
states) on $2\otimes2$ Hilbert space
\begin{align}
\sigma_a = \frac{1}{2} \left(
  \frac{|u\>\<u|}{\|u\|} + \frac{|v_a\>\<v_a|}{\|v_a\|}
\right)
\end{align}
where
\begin{align}
  |u\> &= |00\> + 2|11\> \\
  |v_a\> &= |00\> + a|01\> - 2|10\> - 2|11\>
\end{align}

For the state $\sigma_a$ to have product diagonal in eigenbasis of its
subsystems (which is the necessary condition of $I_M=\Delta^\emptyset$
as shown in previous section) the parameter $a$ must satisfy the
equation
\begin{align}
  \left(4\,a^4 + 40\,a^3 + 87\,a^2 + 160\,a- 341\right) R(a) = 0,
\end{align}
where $R(a)$ is some rational expression.  The polynomial of degree 4
has two real solutions, one around $-8.1$ and one around $1.1$.  Both
give entangled states but the negative one has $I_M-\Delta^\emptyset >
0.09$.  The positive one is
\begin{align}
  a_0 &= {{\sqrt{7200\,\sqrt{z}-\sqrt{w}\,\left(w-504\,z^{{{1}\over{3}}}
        \right)}}\over{8\,w^{{{1}\over{4}}}\,z^{{{1}\over{6}}}}}-{{\sqrt{w}
    }\over{8\,z^{{{1}\over{6}}}}}-{{5}\over{2}} \\
  &\approx 1.10122
\end{align}
where
\begin{align}
  w &= 16\,z^{{{2}\over{3}}}+168\,z^{{{1}\over{3}}}-3111 \\
  z &= {{75\,\sqrt{527523}}\over{16}}-{{131787}\over{64}}
\end{align}

The state $\sigma_{a_0}$ is entangled and satisfies the required
condition of equality and numerical optimizations and parameterized
plots suggest it may fulfill the $I_M=\Delta^\emptyset$ equality.

If we generalize $\sigma_a$ to $\sigma_{a,t}$ where instead of
$|v_a\>$ we take
\begin{align}
  |v_{a,t}\> = |00\> + a|01\> - t|10\> - t|11\>
\end{align}
it seems that for all $t$ starting with some $t_0$ ($1 < t_0 < 2$) we
can find $a_t$ with product diagonal in eigenbasis of subsystems and
$I_M = \Delta^\emptyset$ (or almost equal).  So $\sigma_{a_t,t}$ may
be (for some range of $t$) a class of entangled states satisfying the
equality of $I_M=\Delta^\emptyset$.

\section{Summary}

In our paper we showed that there are states for which quantum
correlations are completely quantum.  We made this by presenting a
family of states for which mutual information (the measure of total
correlations) is equal to quantum deficit zero-way (a measure of
quantumness of correlations).  Surprisingly, the states which we have
found are separable.  We also presented numerical results which
suggest that such situation is possible for entangled states.  What is
intriguing we know that the optimal protocol $\PZ$ which achieves the
value of mutual information is made in local eigenbasis of subsystems
and gives as a result a product state of the form $\varrho_A \otimes
\varrho_B$, where $\varrho_A$ and $\varrho_B$ are states of
subsystems.  This is equivalent to localizing only local information.
Any other local measurement gives us nonproduct classical states, from
which we are able to localize whole its global information but still
less then by only local action.  So the price of producing classical
correlations from which we can get additional information is too high.
The produced entropy is greater than gained information.

\section*{ACKNOWLEDGEMENTS}

We would like to thank Micha\l{} Horodecki and Ryszard Horodecki for
helpful discussions.  This work is supported by EU grant SCALA
FP6-2004-IST no.015714 and by Polish Ministry of Science and Education
under the (solicited) grant No. PBZ-MIN-008/P03/2003.

\bibliography{refbasia}
\end{document}